\documentclass{aa}
\usepackage{graphicx}

\newcommand{\Mpc}{$h^{-1}$\thinspace Mpc}

\def\apj{ApJ}

\begin{document}

\title{Luminosity function and density field of the Sloan and Las
Campanas Redshift Surveys}

\author{G. H\"utsi\inst{1}, J. Einasto\inst{1}, D. L. Tucker\inst{2},
E. Saar\inst{1}, M. Einasto\inst{1}, V. M\"uller\inst{3},
P. Hein\"am\"aki\inst{1,4},  S. S. Allam\inst{5,2}}

\authorrunning{G. H\"utsi  et al.}

\offprints{G. H\"utsi}

\institute{Tartu Observatory, EE-61602 T\~oravere, Estonia \and
Fermi National Accelerator Laboratory, MS 127, PO Box 500, Batavia, IL
60510, USA \and
Astrophysical Institute Potsdam, An der Sternwarte 16, D-14482 Potsdam,
Germany \and
Tuorla Observatory, V\"ais\"al\"antie 20, Piikki\"o, Finland \and
National Research Institute for Astronomy \& Geophysics, Helwan
Observatory, Cairo, Egypt
}

\date{Received 2002 / Accepted ... }

\titlerunning{SDSS and LCRS density field}

\abstract{ We calculate the luminosity function of galaxies of the
Early Data Release of the Sloan Digital Sky Survey (SDSS) and the Las
Campanas Redshift Survey (LCRS). The luminosity function depends on
redshift, density of the environment and is different for the Northern
and Southern slice of SDSS. We use luminosity functions to derive the
number and luminosity density fields of galaxies of the SDSS and LCRS
surveys with a grid size of 1~$h^{-1}$ Mpc for flat cosmological
models with $\Omega _m=0.3$ and $\Omega _\Lambda =0.7.$ We investigate
the properties of these density fields, their dependence on parameters
of the luminosity function and selection effects.  We find that the
luminosity function depends on the distance and the density of the
environment. The last dependence is strong: in high-density regions
brightest galaxies are more luminous than in low-density regions by a
factor up to 5 (1.7 magnitudes).

}

\maketitle

\section{Introduction}

The study of the distribution of matter on large scales is usually
based on the distribution of individual galaxies or clusters of
galaxies. An alternative is to use the density field applying
smoothing of galaxy or cluster distribution with a suitable kernel and
smoothing length. This approach is customary in N-body simulations,
where in each step a smoothed density field is evaluated. To the real
cosmological data the smoothed density method has been applied in the
study of the topology of the galaxy distribution by Gott et
al. (\cite{gmd86}). The IRAS redshift survey was used by Saunders et
al. (\cite{sfr91}) to calculate the density field up to a distance
140~$h^{-1}$\thinspace Mpc. Recently Basilakos et al. (\cite{bpr00})
applied the same method using the PSCz-IRAS redshift survey by
Saunders et al. (\cite{s00}). In all three studies a 3-dimensional
spatial distribution was found. Due to the small volume density of
galaxies with known redshifts a rather large smoothing length was
used. This was sufficient to investigate topological properties of the
galaxy distribution in the first case, and to detect superclusters of
galaxies and voids in other cases.

In this paper, we shall calculate the number and luminosity density
fields based upon the Early Data Release (EDR) of the Sloan Digital
Sky Survey (SDSS) by Stoughton et al. (\cite{s02}) and the Las
Campanas Redshift Survey (LCRS) by Shectman et al. (\cite{Shec96}).
The SDSS Early Data Release consists of two slices of about 2.5
degrees thickness and $65-90$ degrees width, centred on celestial
equator, LCRS consists of six slices of 1.5 degrees thickness and
about 80 degrees width.  The number of galaxies observed per slice
(over 10.000 in the SDSS slices and about 4,000 in LCRS slices) and
their depth (almost 600~$h^{-1}$\thinspace Mpc) are sufficient to
calculate the 2-dimensional density fields with a high resolution
(here $h$ is the Hubble constant in units of
100~km~s$^{-1}$~Mpc$^{-1}$). Using high-resolution number or
luminosity density maps with smoothing scale of the order of 1~$
h^{-1}$\thinspace Mpc\ it is possible to find density enhancements in
the field, which correspond to groups and clusters of galaxies. Using
a larger smoothing length we can extract superclusters of galaxies as
done by Basilakos et al. (\cite{bpr00}). In calculating the density
fields we can take into account most of the known selection effects,
thus we hope that the density field approach gives additional
information on the structure of the Universe on large scales.
Clusters of galaxies from the SDSS were extracted previously by
\cite{2002PASJ...54..515G} using the cut and enhance method.  Loose
groups of galaxies from LCRS were found by Tucker et
al. (\cite{Tucker00}). In accompanying papers by Einasto et
al. (\cite{e02a}, \cite{e02b}, papers II and III, respectively) we use
the density fields of SDSS and LCRS galaxies to derive catalogues of
clusters and superclusters and to compare samples of density-field
defined clusters and superclusters with clusters and superclusters
found with conventional methods.

In the next section we describe the Early Data Release of the SDSS, and the
LCRS samples of galaxies used. In section 3 we derive the luminosity
function for the SDSS and LCRS galaxies using distances found for a
cosmological model with dark matter and energy. In section 4 we calculate
the density fields using SDSS and LCRS galaxy samples. We analyse our
results in section 5, section 6 brings conclusions. 

\section{The data}

\subsection{Early Data Release of the SDSS}

The Early Data Release (EDR) of SDSS consists of two slices about 2.5
degrees thick and $65-90$ degrees wide, centred on celestial equator,
one in the Northern, and the other in the Southern Galactic hemisphere
(Stoughton et al. \cite{s02}), and contains about 35,000 galaxies with
measured redshifts.  SDSS Catalogue Archive Server was used to extract
angular positions, Petrosian magnitudes, and other available data for
all EDR galaxies. From this general sample we obtained the Northern
and Southern slice samples using redshift interval $1000 \leq cz \leq
60000$~km s$^{-1}$, and Petrosian $r^*-$magnitude interval $13.0 \leq r^*
\leq 17.7$.

\subsection{LCRS}

The LCRS is an optically selected galaxy redshift survey which extends to a
redshift of $z\sim 0.2$ and is composed of 6 $1.5^{\circ }\times 80^{\circ }$
slices, 3 in the Northern and 3 in the Southern Galactic hemisphere. The
survey contains 26,418 galaxy redshifts, measured via a 50-fibre or
112-fibre Multi-Object Spectrograph. The 50-fibre fields have nominal
apparent magnitude limits of $16.0\leq R<17.3$, and the 112-fibre fields $%
15.0\leq R<17.7$). A galaxy was included in the calculation of the density
fields if its flux and surface brightness lie within the official survey
limits (see Lin et al. \cite{lks96} or Tucker et al. \cite{Tucker00} for
details). Differences in sampling density and magnitude limits were taken
into account using statistical weighting of galaxies, as will be explained
in section 4.

\section{The luminosity function of the SDSS and LCRS samples}

Luminosity function is one of the basic characteristics of galaxy
population which gives us the co-moving number density of galaxies per
magnitude interval and it is certainly necessary to determine it in
order to understand selection effects in large scale studies. In this
section we are going to estimate luminosity function for the above
mentioned SDSS and LCRS samples.

For the LCRS this analysis has been done for the critical matter
density case by Lin et al. (\cite{lks96}) (L96) but here we are going
to reconsider it in case $\Omega _m=0.3$, $\Omega _\Lambda =0.7.$
Luminosity function for the SDSS galaxies in all filters ($u^*, g^*,
r^*, i^*, z^*$) has
been determined by Blanton et al. (\cite{b01}, hereafter B01) using the
commissioning phase data which covered approximately $230$ square
degrees of the sky along the Celestial Equator in the region bounded
by $145^{\circ }<\alpha <$ $236^{\circ }$ and $
<$ $1.25^{\circ }$. Here we shall calculate it for the SDSS EDR data
which covers $\sim 400$ square degrees of the sky and thus contains
approximately twice as much information.

There exists a variety of methods to estimate luminosity function
e.g. '$1/V_{\max }$ method' by Schmidt (\cite{s68}), parametric
maximum likelihood method by Sandage et al.(\cite{sty79}) (STY79),
'$C^{-}$ method' by Lynden-Bell (\cite{lb71}), stepwise maximum
likelihood method by Efstathiou et al. (\cite{eep88}) etc. In our case
we follow the STY79 approach which assumes special parametric type for
the differential luminosity function $ \phi \left( M\right) $, namely
the one in the form of Schechter function (Schechter \cite{Schechter76})
\begin{eqnarray}
\phi \left( M\right) dM &=&0.4\ln \left( 10\right) \phi ^{*}F\left( M\right)
^{\alpha +1}\exp \left[ -F\left( M\right) \right] dM, \nonumber \\
F\left( M\right) &=&10^{-0.4\left( M-M^{*}\right) },
\end{eqnarray}
and finds maximum likelihood estimates for the free parameters $M^{*}$ and $
\alpha $. Once we have fixed the form for the luminosity function we can
immediately write down the probability that galaxy $i,$ that is observed at
the redshift $z_i,$ has absolute magnitude $M_i$ as 
\begin{eqnarray}
P_i\equiv P\left( M_i\mid z_i\right)& = &\frac{P\left( M_i,z_i\right) }{P\left(
z_i\right) }=P\left( M_i\right) = \nonumber \\
& = & \frac{\phi \left( M_i\right)}{
\int\limits_{M_{\min }\left( z_i\right) }^{M_{\max }\left( z_i\right) }\phi
\left( M\right) dM},
\end{eqnarray}
where the third equality follows from the usual and rather crude assumption
that luminosity function is independent of the spatial location. Here $
M_{\max }\left( z_i\right) $ and $M_{\min }\left( z_i\right) $ determine the
observational window in absolute magnitude at redshift $z_i$ that
corresponds to the surveys limiting apparent magnitudes $m_{\max }$ and $
m_{\min }$, so 
\begin{eqnarray}
\left\{ 
\begin{array}{c}
M_{\min }\left( z_i\right) \\ 
M_{\max }\left( z_i\right)
\end{array}
\right\} =\left\{ 
\begin{array}{c}
m_{\min } \\ 
m_{\max }
\end{array}
\right\}& -&25-5\log d_L\left( z_i\right) -\nonumber \\
&-&k\left( z_i\right) -A\left(l,b\right) ,
\end{eqnarray}
where as usual $d_L\left( z_i\right) $ denotes the luminosity distance, $
k\left( z_i\right) $ is the k-correction and $A\left( l,b\right) $ is the
absorption term. Now we can immediately express the likelihood function $
\mathcal{L}$ for finding $N$ galaxies with absolute magnitudes $M_i$ ($
i=1\ldots N$) as 
\begin{eqnarray}
\mathcal{L} &= & \prod\limits_{i=1}^NP_i\left( \alpha ,M^{*}\right) 
\Rightarrow
\ln \mathcal{L}\left( \alpha ,M^{*}\right) = \nonumber \\
& =& \sum\limits_{i=1}^N\left[ 
\ln\phi \left( M_i\right)
-\ln \int\limits_{M_{\min }\left( z_i\right)}^{M_{\max }\left(
z_i\right) }\phi \left( M\right) dM\right] ,
\end{eqnarray}
and the best fitting values for $M^{*}$ and $\alpha $ are found by
maximising $\ln \mathcal{L}\left( \alpha ,M^{*}\right) $. Assuming
Gaussianity, the corresponding error ellipses are found as usual 
\begin{equation}
\ln \mathcal{L}=\ln \mathcal{L}_{\max }-\frac 12\Delta \chi ^2,
\end{equation}
where $\Delta \chi ^2$ is the critical value for the desired confidence
level for the $\chi ^2$ distribution with two degrees of freedom (Press et
al. \cite{ptv92}). Let us now discuss some pros and cons for the STY79
method.

Pros:
\begin{itemize}
\item[$\bullet$]  as maximum likelihood method it gives us optimal 
estimates 
(i.e. the ones with the minimum variance) for the parameters $M^{*}$ and 
$\alpha ,$

\item[$\bullet$]  STY79 method is insensitive to galaxy clustering and 
density 
evolution
\end{itemize}
... and cons:
\begin{itemize}
\item[$\bullet$] the goodness of fit of the proposed parametric
luminosity function (i.e. Schechter function) cannot be
assessed\footnote{ To overcome this problem Efstathiou et
al. (\cite{eep88}) proposed stepwise maximum likelihood method.},

\item[$\bullet$]  the normalisation $\phi ^{*}$ has to be determined using 
other  methods.
\end{itemize}

The first of these two problems is not really very serious since it is
well known that the Schechter function has almost always served as a
very good approximation and in our case it is important to mention
that earlier studies of the SDSS and LCRS data have certainly
demonstrated this fact (B01, L96). Since our main task here is to
estimate the selection function, which certainly doesn't depend on the
absolute normalisation of the luminosity function, we are going to
determine $\phi ^{*}$ and its uncertainty only for the SDSS data using
rather simple arguments. Namely $ \phi ^{*}$ is obtained so as to
produce the observed number counts and its error is estimated using
re-sampling techniques (to be more precise, ''Jackknife''
method). Here we must also correct for the fact that some galaxies
are missing due to lack of fibres in dense regions, due to
spectroscopic failures as well as fibre collisions. These effects
together give an average sampling rate of 92\% (B01).

In order to calculate the luminosity distance, $d_L(z)
=r( z) ( 1+z) $ (for coordinate distance
$r( z) $ see equation (\ref{eq:cdist}))
as well as to estimate k-corrections, we have to fix the cosmological
model. As already mentioned, we take spatially flat model with
$\Omega _m=0.3$ and $\Omega _\Lambda =0.7$. To estimate k-corrections
\begin{eqnarray}
k(z) &=&2.5\log \left( 1+z\right) +\nonumber \\
& &+2.5\log \left[ \frac{
\int\limits_0^\infty F\left( \lambda \right) R\left( \lambda \right) \lambda
d\lambda }{\int\limits_0^\infty F\left( \frac \lambda {1+z}\right) R\left(
\lambda \right) \lambda d\lambda }\right] ,  \label{eq:kcorr}
\end{eqnarray}
we use spectral templates $F\left( \lambda \right) $ obtained from
Shimasaku (\cite{s98}) and SDSS filter response curves $R\left(
\lambda \right) $ (which include atmospheric and CCD responses) as
compiled by Strauss and Gunn\footnote{ See
http://archive.stsci.edu/sdss/documents/response.dat}. For these
template spectra we first calculate $g^* - r^*$
colours for a range of redshifts. Now, for each observed galaxy, we
interpolate on this grid to find the intrinsic $ g^* 
- r^*$ colours (according to which we classify our
objects) that correspond to the observed colours and redshifts. For
the same template spectra we calculate k-corrections for different
redshifts, and now taking into account the obtained intrinsic
$g^* - r^*$ colours, we can interpolate to find
the desired k-corrections for each galaxy. Finally, we used the
Schlegel et al. (\cite{sfd98}) maps to obtain the reddening-corrected
absolute magnitudes. 

We are going to estimate the luminosity function of the SDSS galaxies only
for the $r^*$ band (i.e. the one with the highest sensitivity)
using the following constraints for the apparent magnitude and recession
velocity:

\begin{itemize}
\item[$\bullet$]  $m_{r^*}: 14.5\ldots 17.5,$

\item[$\bullet$]  $v_{rec}: 1000\ldots 60000$ km/s.
\end{itemize}

Since for the LCRS we have apparent magnitudes only in the ''hybrid''
Kron-Cousins R filter i.e. there is not any colour information, we are
going to approximate (as was also done by L96) spectral energy
distributions by power law $F\left( \nu \right) \propto \nu ^{-\alpha
},$ assuming average spectral index $\alpha \simeq 2.$ The filter is
''hybrid'' since the LCRS photometry was obtained through a Gunn r
filter, although the calibration was done relative to standard stars
in the Kron-Cousins R-band (for more information on LCRS see
e.g. Tucker \cite{t94}). In this case equation (\ref{eq:kcorr}) can be
trivially integrated to give us
\begin{equation}
k(z) =2.5\left( \alpha -1\right) \log \left( 1+z\right) 
\stackrel{\left( \alpha \simeq 2\right) }{\simeq }2.5\log \left( 1+z\right) .
\end{equation}
To estimate Galactic extinction effects for the LCRS magnitudes we use
directly dust maps by Schlegel et al. (\cite{sfd98}) assuming that
these ''hybrid'' magnitudes differ only negligibly from ''true''
Kron-Cousins R magnitudes (see e.g Tucker \cite{t94}) in which case
absorption $A_{R_{KC}}$ is related to the colour excess $E\left( B-V\right) $
through a relation $A_{R_{KC}}=2.673E\left( B-V\right) .$ Magnitude limits
for the LCRS were quoted in the previous section and for the recession
velocity we take an interval $1000\ldots 45000$ km/s. Here, as the
photometry is not so precise as in the case of the SDSS, we take into account
that the observed luminosity function is a convolution of the 'true'
luminosity function with the magnitude errors which we assume to be Gaussian
distributed with the variance $\sigma =0.1$ magnitude (L96). Since in the
LCRS first 20\% of data was obtained using 50-fibre spectrograph and the
rest using 112-fibre system, one must take into account field-to-field
variations in the sampling ratio. Also we correct for the surface brightness
selection effects and for the apparent magnitude and surface brightness
incompletenesses. For more detailed information on these issues we again
refer to L96.

\begin{figure}[tbp]
\centering
\includegraphics[width=0.50\textwidth]{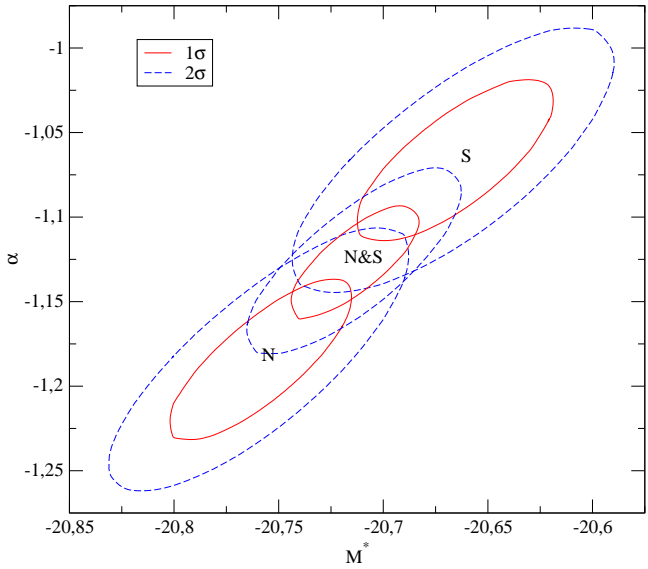}
\caption{$1\sigma$ and $2\sigma$ error ellipses for the SDSS Northern,
Southern and total samples.}
\label{Fig1}
\end{figure}

\begin{figure}[tbp]
\centering
\includegraphics[width=0.50\textwidth]{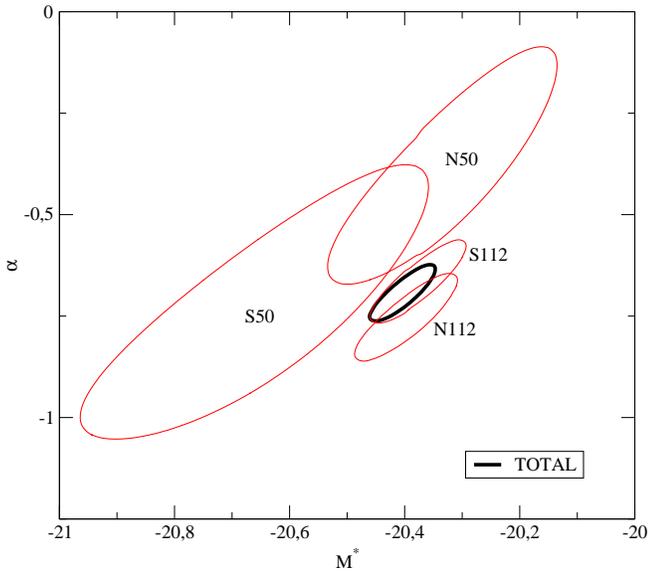}
\caption{$1\sigma$ and $2\sigma$ error ellipses for the LCRS N50, S50, N112,
S112 and total samples.}
\label{Fig2}
\end{figure}

The results for the STY79 maximum likelihood method are given in
Tables \ref{Tab1} and \ref{Tab2} and in Figures \ref{Fig1} and
\ref{Fig2} for the SDSS and LCRS samples, respectively. Slices that
are located in the Northern and Southern Galactic hemispheres are
denoted by N and S, and for the LCRS the numbers 50 and 112 denote the
number of fibres used. For the LCRS this kind of division is the same
as used by L96 and so their Figure 4 (which was obtained assuming
critical matter density) can be compared directly to Figure \ref{Fig2}
(which assumes $\Omega _m=0.3,\ \Omega _\Lambda =0.7$). As already
mentioned, luminosity function normalisation $\phi ^{*}$ was
calculated only for the SDSS data, giving us the value $\phi
^{*}=\left( 151\pm 6\right) \cdot 10^{-4}\ h^3$ Mpc$^{-3}.$ If the
analysis was performed separately for the Northern (N) and Southern
(S) slices, then the values $0.0140\ h^3$ Mpc$^{-3}$ and $0.0158\ h^3$
Mpc$^{-3}$ were obtained for N and S, respectively. If we assumed the
best values for $M^{*}$ and $ \alpha $ obtained for the whole sample
(i.e. $M^{*}=-20.71,\ \alpha =-1.12$) then normalisations for N and S
were $0.0152\ h^3$ Mpc$^{-3}$ and $0.0149\ h^3$ Mpc$^{-3}$,
respectively. Our results for $\alpha $ and $\phi ^{*}$ for the SDSS N
slice (which covers the commissioning phase data analysed by B01)
agree well with the values quoted by B01, although for the Schechter
parameter $M^{*}$ our value $-20.76$ is somewhat lower, agreeing only
marginally with their result.

\begin{figure}[tbp]
\centering
\includegraphics[width=0.50\textwidth]{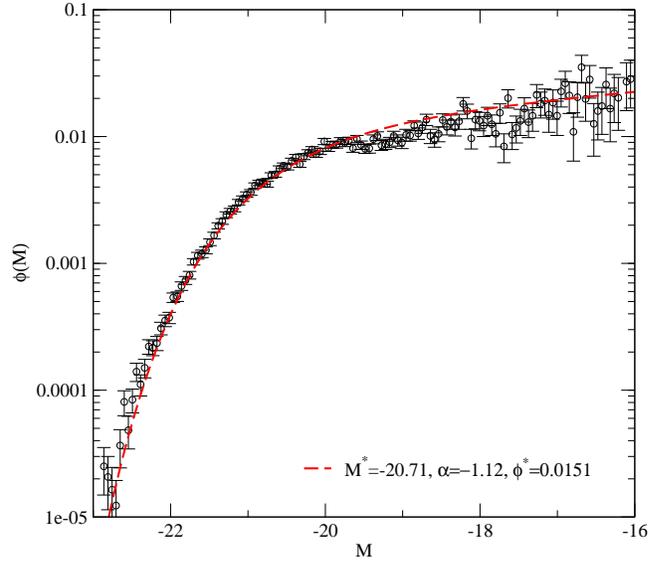}
\caption{Results of the '$1/V_{\max }$ method' for the total SDSS sample
with the corresponding Poissonian error-bars. The best fitting Schechter
function obtained via STY79 maximum likelihood method is also given by a
dashed line.}
\label{Fig3}
\end{figure}

In Figure \ref{Fig3} we compare the best fitting Schechter function obtained
via STY79 method with the results of the simple '$1/V_{\max }$ method' along
with the corresponding Poissonian error-bars. This serves just as an
independent cross-check of our results, although it is of course well known
that '$1/V_{\max }$ method' gives unbiased results only for a homogeneous
distribution (Felten \cite{f76}).

\begin{table}[ht]
\caption{Best fitting $M^{*}$ and $\alpha$ for the SDSS samples}
\label{Tab1}
\begin{tabular}{ccc}
\hline\hline
Sample & $M^{*}-5\log h$ & $\alpha $ \\ \hline
N & $-\left( 20.76\pm 0.04\right) $ & $-\left( 1.19\pm 0.05\right) $ \\ 
S & $-\left( 20.67\pm 0.05\right) $ & $-\left( 1.07\pm 0.05\right) $ \\ 
TOTAL & $-\left( 20.71\pm 0.03\right) $ & $-\left( 1.12\pm 0.03\right) $ \\ 
\hline
\end{tabular}
\end{table}

\begin{table}[ht]
\caption{Best fitting $M^{*}$ and $\alpha$ for the LCRS samples}
\label{Tab2}
\begin{tabular}{ccc}
\hline\hline
Sample & $M^{*}-5\log h$ & $\alpha $ \\ \hline
N50 & $-\left( 20.33\pm 0.12\right) $ & $-\left( 0.40\pm 0.18\right) $ \\ 
S50 & $-\left( 20.64\pm 0.18\right) $ & $-\left( 0.74\pm 0.21\right) $ \\ 
N112 & $-\left( 20.40\pm 0.05\right) $ & $-\left( 0.76\pm 0.07\right) $ \\ 
S112 & $-\left( 20.40\pm 0.05\right) $ & $-\left( 0.70\pm 0.07\right) $ \\ 
NS112 & $-\left( 20.38\pm 0.04\right) $ & $-\left( 0.70\pm 0.04\right) $ \\ 
TOTAL & $-\left( 20.40\pm 0.03\right) $ & $-\left( 0.69\pm 0.04\right) $ \\ 
\hline
\end{tabular}
\end{table}

\section{Density field}

We have extracted from the SDSS and LCRS catalogues subsamples of
galaxies with redshifts $z\leq 0.2$. The calculation of the density
fields consists of three steps: 1) calculation of the distance,
absolute magnitude, and weight factor for each galaxy of the sample;
2) calculation of rectangular coordinates of galaxies and rotation of
coordinate axes in order to minimise projection effects; and 3)
smoothing of the density field using an appropriate kernel and
smoothing length.

Observed redshifts were first corrected for the motion relative to the
CMB dipole (Lineweaver et al. \cite{lts96}). Co-moving distances of
galaxies were calculated as follows (e.g. Peacock \cite{p99}):
\begin{equation}
r(z)=\frac c{H_0}\int\limits_0^z\left\{ 1+\Omega _m\left[ \left( 1+x\right)
^3-1\right] \right\} ^{-\frac 12}dx.  \label{eq:cdist}
\end{equation}
As already mentioned, we used a cosmological model with density
parameters: matter density $\Omega _m=0.3$, dark energy density
$\Omega _\Lambda =0.7$, total density $\Omega _0=\Omega _m+\Omega
_\Lambda =1.0$, all in units of the critical cosmological
density. With these parameters the limiting redshift $z_{lim}=0.2$
corresponds to co-moving distance $r_{lim}=571$~$ h^{-1}$\thinspace
Mpc. Because both surveys cover relatively thin slices on sky, we
shall calculate in the following 2-dimensional density fields,
projecting all galaxies to a plane through the slice.   A 2-dimensional
density field of SDSS EDR was calculated also by Hoyle et
al. (\cite{h02}), who compared the geometry of the large-scale matter
distribution with $\Lambda$CDM simulations.

Our redshift surveys cover a fixed interval in apparent magnitude
which corresponds to a certain range in luminosity, $L_1$ and $L_2$,
depending on the distance of the galaxy. This range, and the observed
luminosity of the galaxy, $L_{obs}$, were found as described in the
previous section.

We regard every galaxy as a visible member of a density enhancement
(group or cluster) within the magnitude window of the group. Further
we suppose that the luminosity function, calculated for the whole
sample, can be applied also for individual groups and clusters.  We
shall discuss in Paper II problems associated with the calculation of
the density field and properties of clusters and superclusters, found
from density field.  Based on these assumptions we calculate the
number and luminosity density fields.  In case of the number density
the weighting factor is proportional to the inverse of the selection
function,
\begin{equation}
w\equiv \frac 1{n^{\mathrm{exp}}(f,D)},  \label{eq:tuck}
\end{equation}
where $n^{\mathrm{exp}}(f,D)$ is the expected number of galaxies in the
whole luminosity range 
\begin{equation}
n^{\mathrm{exp}}(f,D)=F{\frac{\int_0^\infty \phi (L)dL}{\int_{L_1}^{L_2}\phi
(L)dL}}.  \label{eq:dens}
\end{equation}
Here $F$ is the field-to-field sampling fraction, $\phi (L)$ is the
luminosity function, and luminosities $L_1$ and $L_2$ correspond to
the observational window of apparent magnitudes at the distance of the
galaxy. The fraction $F$ takes into account the difference between
50-fibre and 112-fibre data and other effects, for details see L96.
We assume that galaxy luminosities are distributed according to the
Schechter function (Schechter \cite{Schechter76}):
\begin{equation}
\phi (L)\propto (L/L^{*})^\alpha \exp {(-L/L^{*})}d(L/L^{*}),
\label{eq:schechter}
\end{equation}
where $\alpha $ and $L^{*}$ are parameters. The values of these parameters
used in our calculations are given in Table~\ref{Tab3} for each of the
slices. For the LCRS $-6^{\circ }$ N slice we have taken different values
for $\alpha $ and $L^{*}$ since this slice was covered entirely with
50-fibre measurements.

In the case of luminous density every galaxy represents a luminosity 
\begin{equation}
L_{tot}=L_{obs} W_L,  \label{eq:lumin}
\end{equation}
where $L_{obs}=L_{\odot }10^{0.4\times (M_{\odot }-M)}$ is the luminosity of
the visible galaxy of absolute magnitude $M$, $M_{\odot }$ is the absolute
magnitude of the Sun in an appropriate filter. 
In the SDSS $r^*$ filter $M_{\odot }=4.62$ (B01) and in the
Kron-Cousins R-band $M_{\odot }=4.47$ (Binney \& Merrifield \cite{bm98}).
The weight (the ratio of the expected total luminosity to the expected
luminosity in the visibility window) is 
\begin{equation}
W_L =  {\frac{\int_0^\infty L \phi
(L)dL}{\int_{L_1}^{L_2} L \phi (L)dL}}. 
\label{eq:weight}
\end{equation}
We calculated expected luminosities by numerical integration using a
finite total luminosity interval, low limit $L_0$ corresponds to
absolute magnitude $M_0=-13.0$, and upper limit $L_{lim}$ to magnitude
$M_{lim}=-24.5$. In the case of SDSS we adopt the apparent magnitude
window $m_1=17.70$, $ m_2=13.0$ in $r^*$-band; in the case of LCRS the
apparent magnitude window is different for various fields and is taken
into account using corresponding tables.  

This weighting procedure was used also by Tucker et
al. (\cite{Tucker00}) in the calculation of total luminosities of groups
of galaxies.  A different procedure was applied by Moore, Frenk \&
White (\cite{moore93}) (adding the expected luminosity of faint
galaxies outside the observational window).  We assume that density
and luminosity distributions are independent, which leads to a
multiplicative correction.

\begin{table*}[ht]
\caption{Data on SDSS and LCRS slices}
\label{Tab3}
\begin{tabular}{cccccccc}
\hline\hline
Sample & $\delta $ & $\left\langle R.A.\right\rangle $ & $\Delta \delta $ & $
\Delta R.A.$ & $\alpha $ & $M^{*}$ & $N_{gal}$ \\ \hline
SDSS & $0^{\circ }$N & $190.25^{\circ }$ & $2.5^{\circ }$ & $90.5^{\circ }$
& $-1.12$ & $-20.71$ & $15221$ \\ 
SDSS & $0^{\circ }$S & $23.25^{\circ }$ & $2.5^{\circ }$ & $65.5^{\circ }$ & 
$-1.12$ & $-20.71$ & $11864$ \\ 
&  &  &  &  &  &  &  \\ 
LCRS & $-3^{\circ }$N & $191.4^{\circ }$ & $1.5^{\circ }$ & $81.0^{\circ }$
& $-0.69$ & $-20.40$ & $3726$ \\ 
LCRS & $-6^{\circ }$N & $189.8^{\circ }$ & $1.5^{\circ }$ & $77.9^{\circ }$
& $-0.40$ & $-20.33$ & $2132$ \\ 
LCRS & $-12^{\circ }$N & $191.4^{\circ }$ & $1.5^{\circ }$ & $81.1^{\circ }$
& $-0.69$ & $-20.40$ & $3961$ \\ 
LCRS & $-39^{\circ }$S & $12.1^{\circ }$ & $1.5^{\circ }$ & $113.8^{\circ }$
& $-0.69$ & $-20.40$ & $3390$ \\ 
LCRS & $-42^{\circ }$S & $12.2^{\circ }$ & $1.5^{\circ }$ & $112.5^{\circ }$
& $-0.69$ & $-20.40$ & $3610$ \\ 
LCRS & $-45^{\circ }$S & $12.3^{\circ }$ & $1.5^{\circ }$ & $114.1^{\circ }$
& $-0.69$ & $-20.40$ & $3289$ \\ \hline
\end{tabular}
\end{table*}

To find the number and luminosity density fields we calculated for all
galaxies rectangular equatorial coordinates as follows: 
\begin{eqnarray}
x &=&r(z_c)\cos \delta \cos \alpha ,  \nonumber \\
y &=&r(z_c)\cos \delta \sin \alpha , \\
z &=&r(z_c)\sin \delta ,  \nonumber
\end{eqnarray}
here $z_c$ is the redshift, corrected for the motion relative to CMB dipole.
Next we rotated this coordinate system around the $z$-axis to obtain a
situation where the new $y$-axis were oriented toward the average right
ascension of each slice (see Table \ref{Tab3}) and finally we rotated this
new system around its $x$-axis so as to force the average $z$-coordinate to
be zero in order to minimise projection effects.

In Table \ref{Tab3} we also present some of the characteristic
parameters for each of the studied slice as the average declination
$\left\langle \delta \right\rangle $ and right ascension $\left\langle
R.A.\right\rangle$, the number of galaxies, $N_{gal}$, included in the
calculations of the density fields etc.

The EDR of SDSS and LCRS consists of essentially 2-dimensional slices of
about $80^{\circ }$ wide and about 450~$h^{-1}$\thinspace Mpc\ depth. The
width of slices is $1.5-2.5^{\circ }$, thus in space slices form thin
conical (wedge-like) volumes; the thickness of the cones at a
characteristic distance 300~ 
$h^{-1}$\thinspace Mpc\ from the observer is only $8-12$~$h^{-1}$\thinspace
Mpc. Due to the thin shape of slices we shall calculate only a 2-dimensional
density fields.

\begin{figure*}[tbp]
\vspace*{25.0cm}
\caption{Density fields of the SDSS EDR Southern slice, smoothed with $
\sigma= 0.8$~$h^{-1}$\thinspace Mpc\ dispersion. Densities are
expressed in units of the mean density.  Upper panel shows the
number density field, lower panel the luminosity density field.}
\label{Fig4}
\end{figure*}

To smooth the density field two approaches are often used. The first
one uses top-hat smoothing as customary in N-body calculations. In
this case the density in grid corners is found by linear interpolation
where the 'mass' of the particle is divided between grid corners
according to the distance from the respective corner. Variable
smoothing length can be obtained by changing the size of the grid
cell. A better and smoother density field can be obtained when we use
Gaussian smoothing. In this case the 'mass' of the particle is
distributed between neighbouring cells using the Gaussian smoothing
\begin{equation}
W_{jg}=\left( 2\pi \sigma _{\mathrm{sm}}^2\right) ^{-\frac 12}\ \exp \left( -
\frac{|\mathbf{x}_j-\mathbf{x_g}|^2}{2\sigma _{\mathrm{sm}}^2}\right) ,
\label{eq:gauss}
\end{equation}
here $\mathbf{x}_j$ and $\mathbf{x_g}$ are positions of the galaxy and the
grid cell, respectively, and $\sigma _{\mathrm{sm}}$ is the smoothing length.

To calculate the density fields, we have formed a grid of cell size
1~$ h^{-1} $\thinspace Mpc. On farther part of the survey only very
bright galaxies can be observed, the weight factor becomes large and
the fields are noisy. To decrease this effect we have calculated the
density fields for the LCRS only for co-moving distance up to
$r_{lim}\leq 450$~$h^{-1}$\thinspace Mpc. Very close to the observer
the density fields are also noisy. Here slices become very thin in
real space, and only galaxies of low luminosity fall into the window
of apparent magnitudes observed. For nearby region the density fields
were calculated but density enhancements were not considered as
clusters if $r\leq 50$~$h^{-1}$\thinspace Mpc.

In calculations we have used several smoothing lengths. First, we tried a
smoothing length $\sigma _{\mathrm{sm}}=2$~$h^{-1}$\thinspace Mpc. This
smoothing yields a very smooth field where density enhancements (clusters
and groups of galaxies) can be easily identified. However, the size of these
enhancements clearly exceeds the size of real clusters of galaxies. Thus we
have used a smaller smoothing length, $\sigma _{\mathrm{sm}}=0.8$~$h^{-1}$
\thinspace Mpc, to calculate a density field which can be considered as an
approximation to the true density field of dark matter associated with
luminous galaxies.

To identify superclusters of galaxies we have used a larger smoothing
length, $\sigma _{\mathrm{sm}}=10$~$h^{-1}$\thinspace Mpc. Experience
with numerical simulations has shown that this smoothing length is
suitable to select supercluster-size density enhancements (Frisch et
al. \cite{f95}). To take into account the thickness of the slice, the
smoothed density fields were divided by the volume of the column at
the location of the particular cell in the real 3-D space, covered by
observations of the slice. In this way the density fields are reduced
to a plane parallel sheet of constant thickness.

The high-resolution number and luminosity density fields for the SDSS
EDR Southern slice are shown in Figure~\ref{Fig4}. The 
density fields of all EDR SDSS slices are presented
in Einasto et al. (\cite{e02a}), the fields of LCRS slices are
presented in Einasto et al. (\cite{e02b}).

\section{Analysis}

So far we have determined the number and luminosity density fields for the
SDSS EDR and LCRS samples. In order to estimate the selection effects we
have calculated the luminosity function assuming Schechter type of
parametrisation.

\begin{figure}[tbp]
\vspace{10.8cm}
\caption{Upper and lower panels show respectively the number and
luminosity density as a function of coordinate
distance for the SDSS Northern and Southern slices using the best
fitting Schechter parameters for the total sample as given in Table
\ref{Tab1}.  Densities were calculated using the high-resolution
field.}
\includegraphics{hytsi_fig5a.eps}
\includegraphics{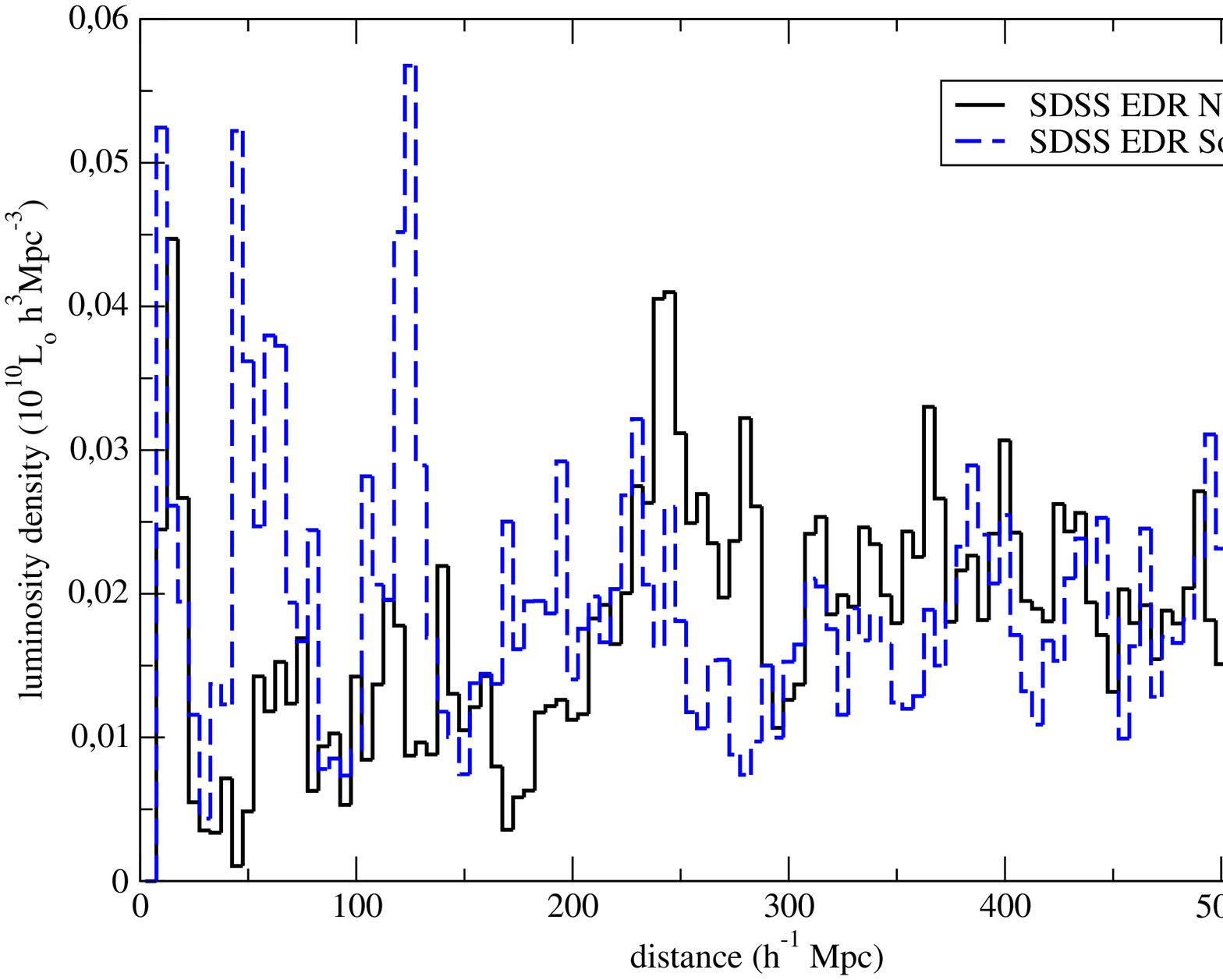}
\label{Fig5}
\end{figure}

To be sure that we are not grossly under- or overestimating the
statistical weights we are applying to  visible objects (in order to
take account their invisible companions), we perform some simple
calculations. Namely, we calculate the number and luminosity density
averaged over thin shells as a function of the coordinate
distance. What concerns the number density, then we expect it to fall
in the distant part of the samples since there we miss some of the
faintest systems totally. The same need not be true for the
luminosity density since even the modest amount of luminosity
evolution could compensate the decrease in the number density. The
results for the number and luminosity density for the SDSS EDR data
are given in Figure \ref{Fig5}. We see that indeed the number density
starts to decline at the largest distances, as expected, indicating
that we are at the right track.  Contrary to the number density the
luminosity density is almost constant for the Northern slice, but
shows a slightly increasing trend in the Southern slice on large
distances. This is also evident from the Figure \ref{Fig4} where
densities are colour coded.  For the LCRS slices we get essentially
the same results and for the brevity we are not going to present them
here. All this suggests that our statistical weighting scheme looks
quite reasonable for such a preliminary study.

\begin{figure}[tbp]
\centering
\includegraphics[width=0.50\textwidth]{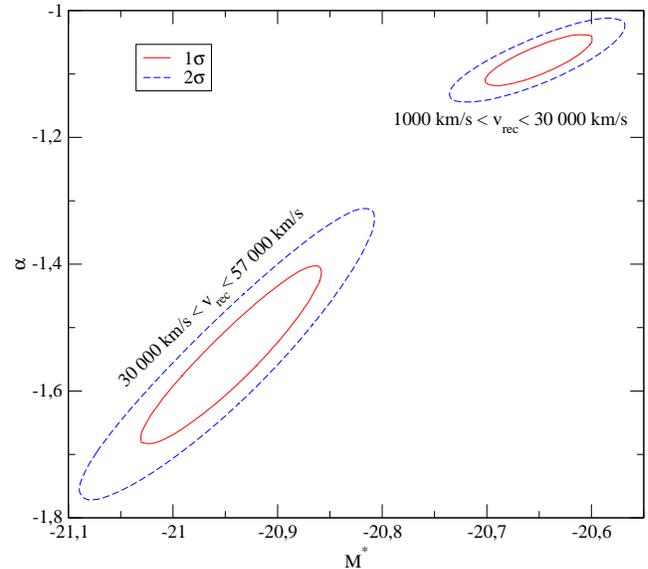}
\caption{$1\sigma$ and $2\sigma$ error ellipses for the SDSS data divided
into low $z$ and high $z$ subsamples.}
\label{Fig6}
\end{figure}

\begin{figure}[tbp]
\centering
\includegraphics[width=0.50\textwidth]{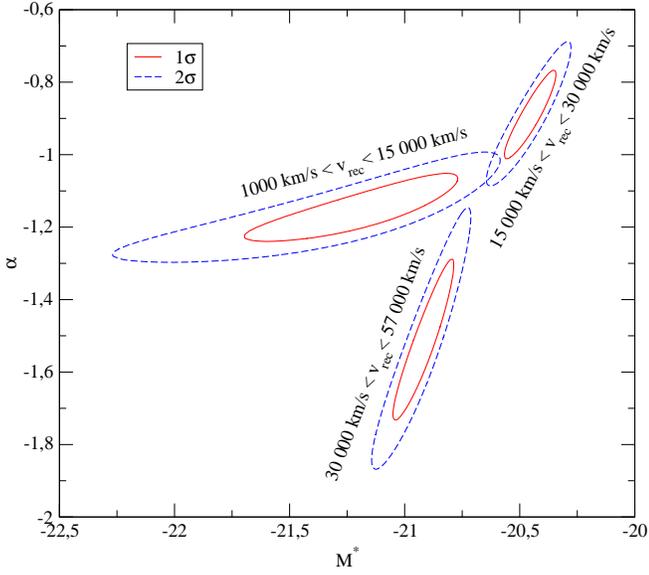}
\caption{$1\sigma$ and $2\sigma$ error ellipses for the SDSS EDR Southern
slice data divided into three $z$ intervals.}
\label{Fig7}
\end{figure}

The second thing we want to mention is the fact that we have assumed
that the same Schechter parameters $M^{*}$ and $\alpha $ apply for the
whole sample. In order to study the possible evolution with redshift
$z$ we have divided our SDSS EDR sample into two parts: low $z$ part
($v_{rec}=1000\ldots 30000$ km/s) and high $z$ part
($v_{rec}=30000\ldots 57000 $ km/s) and performed the STY79 maximum
likelihood analysis for both of these separately. The results that are
given in Figure \ref{Fig6} probably suggest that $M^{*}$ and $\alpha $
are evolving in time. From this figure we obtain a simple linear
approximation
\begin{eqnarray}
M^{*} &=&-4.2z-20.37,  \nonumber  \label{eq:last} \\
\alpha &=&-6.7z-0.63,
\end{eqnarray}
where we have taken into account the fact that the average $z$ for the
low $z$ sample is 0.067 and for the high $z$ one 0.136. If we assume
this kind of $M^{*}$ and $\alpha $ dependences on $z$ for the whole
sample and calculate the number and luminosity densities as above then
we get pretty bad results. So certainly this kind of big linear
variation with $z$ makes things worse.

\begin{figure}[tbp]
\centering
\includegraphics[width=0.50\textwidth]{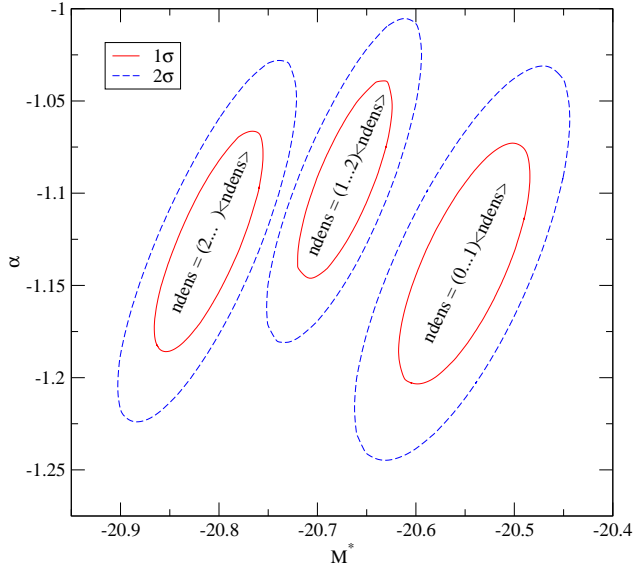}
\caption{$1\sigma$ and $2\sigma$ error ellipses for the SDSS EDR Southern
slice data divided into three global density $\delta$ intervals.}
\label{Fig8}
\end{figure}

\begin{table}[ht]
\caption{Luminosity function parameters in different environment}
\label{Tab4}
\begin{tabular}{ccc}
\hline\hline
Sample & $M^{*}-5\log h$ & $\alpha $   
\\ \hline
\\
$0 < \delta_0 \leq 1.0$   & $-20.55 \pm 0.07$ & $ -1.14 \pm 0.06$  \\
$1.0 < \delta_0 \leq 2.0$ & $-20.67 \pm 0.07$ & $-1.09 \pm 0.05$ \\
$2.0 < \delta_0         $ & $-20.81 \pm 0.07$ & $-1.13 \pm 0.05$ \\
\\ \hline
\end{tabular}
\end{table}

Next we divide the SDSS EDR Southern sample into three parts:
$v_{rec}=1000\ldots 15000$ km/s, $v_{rec}=15000\ldots 30000$ km/s and
$ v_{rec}=30000\ldots 60000$ km/s. In this case we get non-monotonic
dependences for $M^{*}$ and $\alpha $ on $z$ (see Figure \ref{Fig7}).
We also notice that variations in $M^{*}$ and $\alpha $ are correlated
with the density of the surroundings: due to the effects of the nearby
dense regions the value for $M^{*}$ is smaller than on average; as we
move further away $M^{*}$ gets bigger, which is caused by the large void
at distances $\sim 200\ldots 350\ h^{-1}$ Mpc (see Figure
\ref{Fig4}). In distant parts, as the density rises again, $M^{*}$
gets accordingly smaller. This suggests that environmental effects are
more important than evolutionary ones (which we expect to be monotonic
in $z$), and so the usual assumption in determining the luminosity
function $\phi \left( M\right) $ that it is independent of the spatial
position cannot be strictly correct.

\begin{figure}[tbp]
\vspace*{8.0cm}
\caption{Luminosities of Southern slice galaxies as a function of
the global relative density $\delta$ smoothed with 2~\Mpc\ dispersion.
Luminosities are expressed in units of $10^{10}$ Solar luminosities,
global density $\varrho$ in units of the mean density.  The Northern
slice has similar dependence on the environmental density.}
\includegraphics{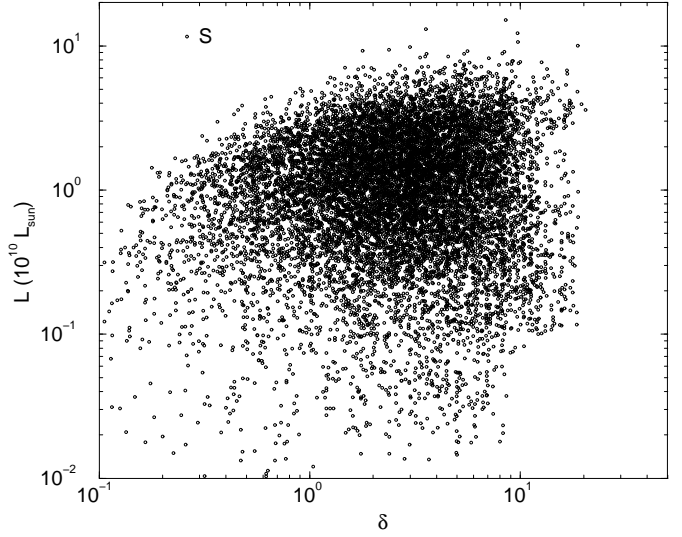}
\label{Fig9}
\end{figure}

To study this aspect in more detail we calculated the dependence of
luminosity function parameters on the density of the environment.  We
calculated for every galaxy the global density $\delta_0$ (with
10~\Mpc\ smoothing) and used this value as environmental parameter.
Galaxies were divided into three types according to density (expressed
in units of the mean density).  Parameters of the luminosity function
are given in Table~\ref{Tab4} and Figure~\ref{Fig8}.  We see that
parameter $M^{*}$ changes considerably with density: in high-density
regions its value is more negative, i.e. galaxies are brighter.  A
similar tendency has been recently observed in other surveys
(\cite{1998ApJ...505...25B}, \cite{2000ApJ...545....6B},
\cite{2001MNRAS.328...64N}).

The density field with 10~\Mpc\ smoothing characterizes the density on
supercluster scales.  The dependence of galaxy luminosity on density
has been found also on smaller scales.  In Figure~\ref{Fig9} we plot
luminosities of galaxies of the Southern SDSS EDR slice as a function
of the density smoothed on 2~\Mpc\ scale.  This smoothing is sensitive
to density changes within clusters of galaxies and in galaxy
filaments.  We see a strong dependence of the upper end of
luminosities: brightest galaxies in low-density regions are about 5
times less luminous than in high-density regions. This difference in
luminosity corresponds to a difference 1.7 magnitudes, much more than
expected from the analysis based on larger smoothing length.  We plan
to address all these issues in more detail in the forthcoming papers.
Physical interpretation of these phenomena needs theoretical studies.

We repeat that all the density fields (see Figure \ref{Fig4}) compiled
in this paper refer to the redshift space.

\section{Conclusions}

In this paper we have calculated the galaxy luminosity function for
the SDSS EDR and LCRS samples and used it to construct the number and
luminosity density fields (smoothed on $0.8\ h^{-1}$ Mpc scale)
assuming flat underlying cosmologies with $\Omega _m=0.3$ and $\Omega
_\Lambda =0.7.$ The analysis presented here is rather simple and
serves as a first step in the study of the distribution of galaxies in
SDSS and LCRS samples.  The principal conclusion from our study is:
parameters of the galaxy luminosity function depend on the distance
from the observer, density of the environment, they are different for
the Northern and Southern slice.  The largest effect
is the dependence on the density of the environment: in high-density
regions brightest galaxies are more luminous than in low-density
regions by a factor up to 5 (1.7 magnitudes).  Some of these effects
suggest that it is not yet possible to find an universal set of
parameters of the luminosity function valid for a fair sample of the
Universe.  In other words, presently available samples are still too
small to be considered as candidates for the fair sample.

\begin{acknowledgements}

The present study was supported by Estonian Science grants ETF 3601,
ETF 4695 and TO 0060058S98.  P.H. was supported by the Finnish Academy of
Sciences.  J.E. thanks Fermi-lab and Astrophysikalisches Institut
Potsdam for hospitality where part of this study was performed.
\\
Funding for the creation and distribution of the SDSS Archive has been
provided by the Alfred P. Sloan Foundation, the Participating
Institutions, the National Aeronautics and Space Administration, the
National Science Foundation, the U.S. Department of Energy, the
Japanese Monbukagakusho, and the Max Planck Society. The SDSS Web site
is http://www.sdss.org/.
\\
The SDSS is managed by the Astrophysical Research Consortium (ARC) for
the Participating Institutions. The Participating Institutions are The
University of Chicago, Fermi-lab, the Institute for Advanced Study, the
Japan Participation Group, The Johns Hopkins University, Los Alamos
National Laboratory, the Max-Planck-Institute for Astronomy (MPIA),
the Max-Planck-Institute for Astrophysics (MPA), New Mexico State
University, University of Pittsburgh, Princeton University, the United
States Naval Observatory, and the University of Washington.

\end{acknowledgements}

\end{document}